\documentclass[aps,pre,reprint]{revtex4-1}

\usepackage{amsmath,amssymb,amsfonts,amsthm}
\usepackage{bm,graphicx}

\newcommand{\avg}[1]{\langle#1\rangle}
\newcommand{\pnt}[1]{{\bm #1}}

\newcommand{\oper}[1]{\mathcal{#1}}

\DeclareMathOperator{\const}{\mathrm{const}}
\DeclareMathOperator{\imi}{\mathrm{i}}
\newcommand{\Sec}[1]{\textit{#1}}

\begin{document}
\footnotetext[0]{See Supplemental Material}

\title{
  A Volterra-series approach to stochastic nonlinear dynamics: The Duffing oscillator
  driven by white noise
}
\date{\today / Revision 4}
\author{Roman Belousov}\email{belousov.roman@gmail.com}
\author{Florian Berger}
\author{A. J. Hudspeth}
\affiliation{Howard Hughes Medical Institute, Laboratory of Sensory Neuroscience, The Rockefeller University, New York, NY 10065, USA}
\begin{abstract}
  The Duffing oscillator is a paradigm of bistable oscillatory motion in physics,
  engineering, and biology. Time series of such oscillations are often observed experimentally
  in a nonlinear system excited by a spontaneously fluctuating force. One is then
  interested in estimating effective parameter values of the stochastic Duffing model
  from these observations---a task that has not yielded to simple means of analysis.
  To this end we derive theoretical formulas for the statistics of the Duffing oscillator's
  time series. Expanding on our analytical results, we introduce methods of statistical
  inference for the parameter values of the stochastic Duffing model. By applying
  our method to time series from stochastic simulations, we accurately reconstruct
  the underlying Duffing oscillator. This approach is quite straightforward---similar
  techniques are used with linear Langevin models---and can be applied to time series
  of bistable oscillations that are frequently observed in experiments.
\end{abstract}
\maketitle

Some of the most interesting and complex behaviors in nature emerge from coupled systems
with nonlinearities embedded in an environment. Depending on the relevant time and
length scales, influences from the environment can be described effectively as fluctuating
forces driving such systems. A fundamental example of such a system is the stochastic
Duffing oscillator, which, together with its generalizations, has various applications
in engineering and biophysics \cite{Khan2001,Chatterjee2003,Alonso2014,Mindlin2017,Cherevko2016,Parshin2016,Cherevko2017,FitzHughNagumo}.
The Duffing equation offers the simplest nonlinear model that describes bistable
oscillatory motion \cite[Sec. 7.6]{StrogatzII}. Under certain physical conditions
the equation represents a power-series approximation for a general class of Lienard
systems \cite[Sec. 7.4]{StrogatzII}.

The Duffing model extends the harmonic oscillator by adding a cubic nonlinear term:
\begin{equation}\label{eq:main}
  \ddot{x} + a \dot{x} + b x + c x^3 = f
\end{equation}
for an unknown function of time $x(t)$ and an external force $f(t)$. The constants
$a$, $b$, and $c$ are the damping coefficient, the linear stiffness, and the cubic
Duffing parameter, respectively. Because the above equation is of second order in
time, the phase of this system is specified by two degrees of freedom $(x,\dot{x})$.

In various situations the form of the relevant driving force is $f(t) = A \dot{w}(t)$,
in which $A > 0$ is a constant and $\dot{w}(t)$ is Gaussian white noise of zero mean
and unit intensity. Equation~(\ref{eq:main}) describes a stable dynamical system when
the coefficients $a > 0$ and $c > 0$ are strictly positive. Unlike the harmonic oscillator,
for which $c = 0$, the Duffing model admits a negative linear stiffness $b$.

The Duffing oscillator is bistable when $b < 0$. Its phase space is symmetric about
the origin $(x,\dot{x}) = (0,0)$, which represents an unstable fixed point in absence
of external force. Two stable equilibria occur at $(\xi,0)$ and $(-\xi,0)$, in which
$x=\pm\xi=\pm\sqrt{-b/c}$ correspond to the minima of the Duffing double-well
potential $U(x) = \const + b x^2/2 + c x^4/4$. In the monostable regime, for which
$b \ge 0$, the origin is the only fixed point.

A problem that arises often in quantitative studies of bistable nonlinear systems
is the determination of a model's parameter values. In experiments one usually observes
time series of noisy oscillations. The model parameters may then be adjusted empirically
to reproduce the measurements as closely as possible. This method is rather arbitrary
and imprecise, whereas other available approaches require additional experimental
data \cite{Chatterjee2010,Chatterjee2003,Smelyanskiy2005,HeQie2007,Quaranta2010}.

Although a time series of oscillations may in principle contain enough information to infer
the parameter values of the Duffing oscillator, this approach has not been duly pursued.
In the present letter we derive statistical formulas for the time series $x(t)$ in
the regime of bistable oscillations. These expressions rely on the Volterra expansion
of functionals \cite[Chapters 1-3]{Rugh}, which provide the mathematical framework
of nonlinear response theory \cite{Peterson}. Expanding on our analytical results,
we then develop statistical methods to estimate the parameter values of the stochastic
Duffing Eq.~(\ref{eq:main}) from the time series $x(t)$.

\Sec{General theory}.---The functional series of Volterra generalize the Taylor-Maclaurin
expansion of functions in calculus \cite[Sec. 1.5]{Rugh}. In particular, we can represent
the solution of Eq.~(\ref{eq:main}) as a functional of the force $f(t)$:
\begin{multline}\label{eq:Volterra}
  x(t|f) = x_0(t) + \int_0^t dt_1\, g_1(t-t_1) f(t_1)\\
    + \iint_0^t dt_1 dt_2\, g_2(t-t_1,t-t_2) f(t_1) f(t_2) + ...\\
    = x_0(t) + \gamma_1(t) + \gamma_2(t) + ...
\end{multline}
Here $g_1$ and $g_2$ are the Volterra kernels of the linear and quadratic terms in
$f$, $\gamma_1(t)$ and $\gamma_2(t)$, respectively.

Provided that the series~(\ref{eq:Volterra}) converge, a truncated Volterra expansion
approximates the solutions of Eq.~(\ref{eq:main}). We find the unknown kernels $g_{i=1,2...}$
by using the variational approach \cite[Sec. 3.4]{Rugh}: we replace the external
force $f(t)$ by a constant $f_c \equiv \const$ and substitute Eq.~(\ref{eq:Volterra})
into (\ref{eq:main}). Then, by collecting terms with coefficients of equal powers
in $f_c$, we obtain the following system of equations
\begin{eqnarray}
  \label{eq:zero}
  	0 &=& \ddot{x}_0 + a \dot{x}_0 + b x_0 + c x_0^3,
    \\\label{eq:one}
  	f &=& \ddot\gamma_1 + a \dot\gamma_1 + b \gamma_1 + 3 c x_0^2 \gamma_1,
    \\\label{eq:two}
  	0 &=& \ddot\gamma_2 + a \dot\gamma_2 + b \gamma_2 + 3 c (x_0 \gamma_1^2 + x_0^2 \gamma_2),\\
      &...&\nonumber
\end{eqnarray}
Equation~(\ref{eq:zero}), which defines $x_0(t)$, is equivalent to the homogeneous
Duffing problem (\ref{eq:main}) with $f\equiv0$. The Volterra kernels can be found
in successively increasing orders from the linear Eqs.~(\ref{eq:one}), (\ref{eq:two}),
\textit{etc.}

The equilibrium solution $x_0(t)\equiv0$ of Eq.~(\ref{eq:zero}) spawns a particularly
convenient set of Eqs.~(\ref{eq:one}) and (\ref{eq:two}) for the monostable Duffing
oscillator \cite{Khan2001}. In the bistable case, the kernels of the Volterra series
at $x_0(t)\equiv0$ diverge with $t\to\infty$ \cite[Secs. \ref{sec:mono} and \ref{sec:bi}]{Note0}. In fact,
this expansion may even fail to exist \cite{Ku1966}. Therefore we develop the Volterra
series at the stable equilibria $x_0(t)\equiv\pm\xi$.

As a generalization of the Taylor-Maclaurin series, the Volterra expansion may be
limited by a convergence region. Moreover the accuracy of the truncated expression
deteriorates as $f(t)$ becomes progressively greater: the relevant physical scales
are introduced later. Due to the symmetry of the Duffing oscillator, $x_0(t)\equiv\xi$
and $x_0(t)\equiv-\xi$ lead to identical odd-order terms in Eq.~(\ref{eq:Volterra}),
whereas the even-order terms differ by a factor of $-1$ \cite[Sec. \ref{sec:bi}]{Note0}.
As we show shortly, these two series are accurate in the neighborhood of the expansion
points as long as the system's trajectory $x(t)$ does not cross the special point
$x=0$.

If the amplitude $A$ of the external force $f(t)$ is small, the Duffing oscillator
remains in one of the two potential wells at $x=\pm\xi$. The truncated Volterra expansions
then describe the solutions of Eq.~(\ref{eq:main}) accurately around the respective
equilibrium points. The linear response of $x(t)$ is harmonic in the first order
of the parameter $A$,
\begin{equation}\label{eq:hrm}
  \ddot{\gamma}_1 + a \gamma_1 - 2 b \gamma_1 = f,
\end{equation}
which can also be obtained by linearization of Eq.~(\ref{eq:main}) at the minima
of the Duffing potential.

When $A$ is sufficiently large, the Duffing oscillator undergoes stochastic transitions
between the two potential wells. The statistical average $\avg{x}=0$ vanishes due
to the symmetry of the problem. Although truncated Volterra expansions of $x(t)$
are inaccurate in this case, Eq.~(\ref{eq:Volterra}) may still be applied to describe
pieces of the oscillator's trajectory in an $\varepsilon$-neighborhood of each potential
well ($|x(t)\pm\xi| \le \varepsilon < \xi$). A physical assumption is implied thereby
that the external force does not perturb the system's energy much while the oscillator
remains in one of the wells. In this sense the argument $f$ of the functional Eq.~(\ref{eq:Volterra})
is small. Statistically the selected pieces of the oscillator's trajectory belong to two ensembles of \textit{conditional probability distributions}
$p(x_\xi) = p(x|\,|x-\xi|\le\varepsilon)$ and $p(x_{-\xi}) = p(x|\,|x+\xi|\le\varepsilon)$
\cite{{[{A similar approach has already been applied in }] [{, Sec. 4 to describe fluctuations in a metastable dynamical system.}]Belousov2014}}.

The energy barrier that separates the two wells of the Duffing potential becomes
negligible for external forces of extreme amplitudes $A$. The oscillations then resemble
those of a monostable regime.

\Sec{Statistical analysis}.---The time-invariant probability density of a bistable
Duffing oscillator driven by white noise is generally bimodal. Two Gaussian-like
peaks correspond to the minima of the double-well potential $U(x)$, for which the
harmonic oscillator Eq.~(\ref{eq:hrm}) describes the local dynamics of $x(t)$. We
can construct the time-invariant probability density of $x$ by using an exponential
form:
\begin{equation}\label{eq:poly}
  p(x) \approx Z^{-1} \exp[-P(x) + \mathcal{O}(x^5)],
\end{equation}
in which $Z$ is a normalization constant and $P(x)\ge0$ is a polynomial of fourth
order in $x$ \cite{Belousov2016,Belousov2014}. At the two global minima of Eq.~(\ref{eq:poly})
the following conditions must be satisfied:
$$ P(\pm\xi) = 0,\quad \frac{\partial^2 P}{\partial x^2}(\pm\xi) = -4 a b / A^2. $$
The last equality ensures that the Laplace approximation of $p(x)$ at $x=\pm\xi$
\cite{Touchette} obeys the statistics of Eq.~(\ref{eq:hrm}) \cite[Sec. II-3]{Chandrasekhar}.
Owing to the symmetry of the bistable Duffing system $p(x)=p(-x)$, the general form
of $P(x)$ is given by
$$ P(x) = \frac{a c (x^2-\xi^2)^2}{2 A^2}, $$
which leads to
\begin{equation}\label{eq:px}
  p(x) \approx Z^{-1} \exp\left[-\frac{a c (x^2-\xi^2)^2}{2 A^2}\right]
    \propto \exp\left[- \frac{2 a U(x)}{A^2} \right].
\end{equation}
The normalization constant $Z$ can be found by integration of the exponential factor
in the above equation:
$$
  Z = \frac{\pi\xi}{2} \exp(-z) [I_{-1/4}(z) + I_{1/4}(z)],
$$
in which $z=a c \xi^4/(4 A^2)$ and $I_i(\cdot)$ is the $i$\textsuperscript{th}-order modified Bessel
function of the first kind.

In addition the autocorrelation function $\chi(t)$ can be approximately calculated
for $x_{\pm\xi}(t)$ from Eq.~(\ref{eq:hrm}) \cite{BelousovCohen}:
\begin{multline}\label{eq:crr1}
  \chi(t) \approx \chi_1(t) = \frac{\avg{\gamma_1(0)\gamma_1(t)}}{\avg{\gamma_1}^2}
    \\ = \exp\left( -\frac{a\,|t|}{2} \right) \left[\cos(\Omega\,|t|) + \frac{a}{2\Omega}\sin(\Omega\,|t|)\right],
\end{multline}
in which $\Omega = \sqrt{-2 b - a^2/4}$ \footnote{
  If $\Omega^2 < 0$ one should use its absolute value instead and replace the trigonometric
  functions of cosine and sine in Eq.~(\ref{eq:crr1}) by the hyperbolic ones \cite[Sec. II-3]{BelousovCohen,Chandrasekhar}.
}.

\Sec{Dimensional analysis}.---Equations~(\ref{eq:main}), (\ref{eq:px}), and (\ref{eq:crr1})
characterize physical scales of the Duffing oscillator. First we adopt the constants
$a^{-1}$ and $a/\sqrt{c}$ as the units of $t$ and $x$ respectively. The energy scales
are then determined by the height of the barrier between the wells of the Duffing
potential $\epsilon = b^2 / (4 c)$. It suffices therefore to consider $b = -2 \sqrt{\epsilon c}$
as a typical value for a unit energy barrier $\epsilon$.

The Boltzmann-like factor $\exp[-2 a U(x)/A^2]$ in Eq.~(\ref{eq:px}) relates the
level of  energy fluctuations $A^2/(2 a)$ in the system to the Duffing potential
$U(x)$. This helps us identify the amplitudes of the external force $A \lesssim \sqrt{2 a \epsilon}$,
for which the truncated Volterra series might be useful to describe the trajectories
$x_{\pm\xi}(t)$.

Because Eq.~(\ref{eq:crr1}) was derived from the linear-response approximation, it
is independent of $A$. Although this formula is very convenient, its accuracy is
limited to small time and energy scales, as shown below. The autocorrelation function
Eq.~(\ref{eq:crr1}) decays exponentially with a relaxation time $\tau \approx a^{-1}$
\cite{Zwanzig1969}. In general we may expect the formula~(\ref{eq:crr1}) to hold
for $0 \le t \lesssim \tau$.

\Sec{Parametric inference}.---To test our theoretical results we simulated Eq.~(\ref{eq:main})
\cite[Sec. \ref{sec:sim}]{Note0} by using an operator-splitting algorithm \cite[Appendix C]{Tuckerman1992,Belousov2017}.
The results are reported in the system of units reduced by the time, length, and energy
constants $a^{-1}$, $a/\sqrt{c}$, and $\epsilon$, respectively. As justified earlier,
the constant $b = -2$ is fixed. Our simulations differ only by values of the parameter
$A$.

\begin{figure}
\includegraphics[width=1\columnwidth]{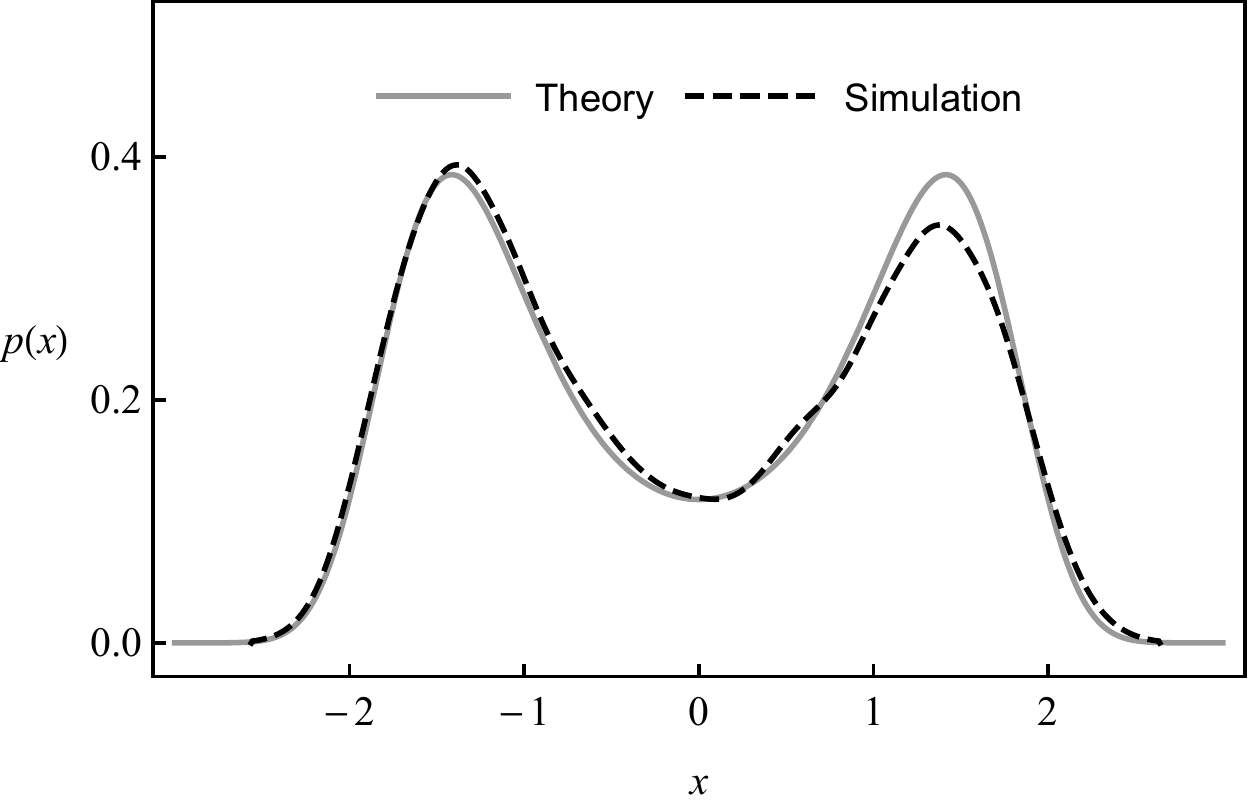}
\caption{\label{fig:px}
  Steady-state probability density $p(x)$ for the time series simulated with $A=1.3$.
  The theoretical expression is given by Eq.~(\ref{eq:px}), whereas the computational
  results are represented by a smooth histogram \cite{Mathematica}.
}
\end{figure}

\begin{figure}
\includegraphics[width=1\columnwidth]{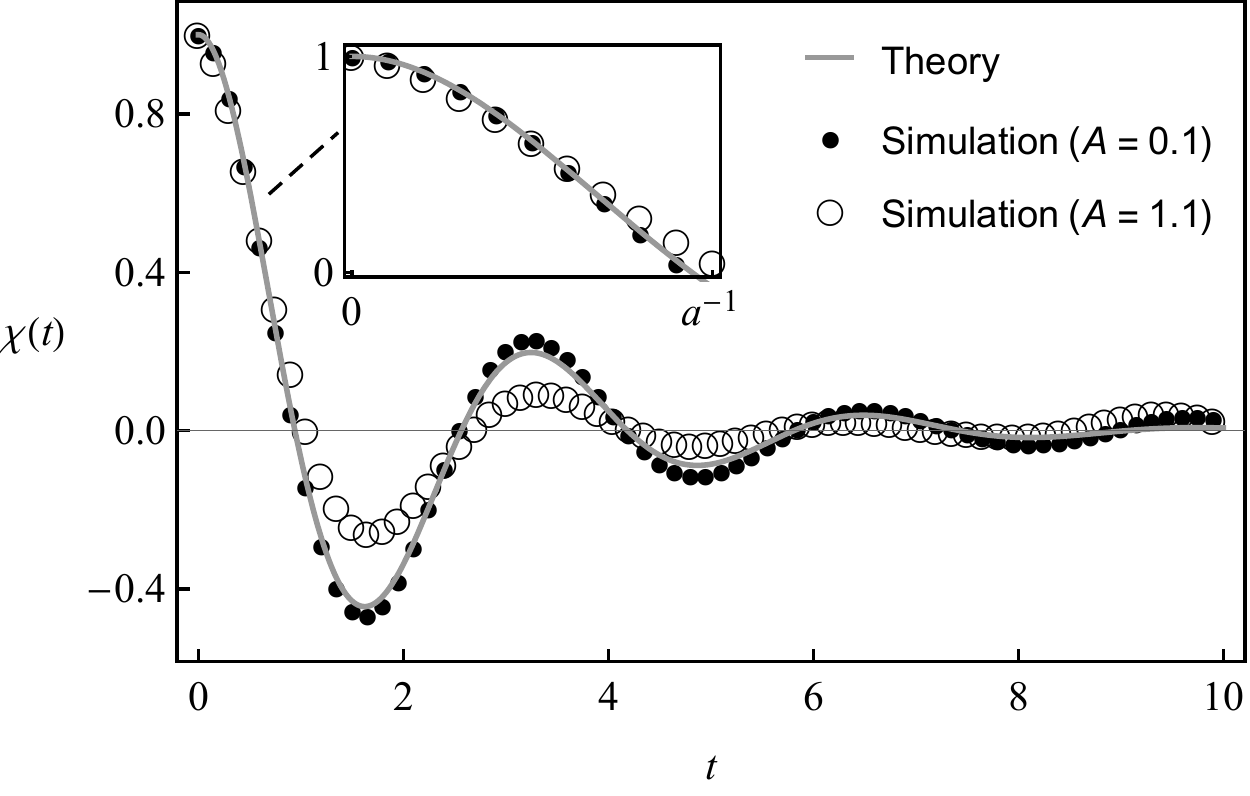}
\caption{\label{fig:crr1}
  Time autocorrelation function $\chi(t)$ for time series $x(t)$ driven by a weak
  external force ($A=0.1$) and for the time series $x_\varepsilon(t)$ ($A=1.1$).
  Error bars, which are comparable in size to the plot markers, are omitted. The
  theoretical curve corresponds to Eq.~(\ref{eq:crr1}). The inset magnifies the exponential
  decay for $t \lesssim a^{-1}$.
}
\end{figure}

\begin{table}[b]
\caption{\label{tbl:fit}
  Statistical inference of parameter values for the Duffing oscillator Eq.~(\ref{eq:main}),
  which was simulated with fixed values $a=1$, $b=-2$, and $c=1$; $A$ varied in the
  range $[0.6,1.3]$. The estimated parameter values are denoted by $\hat{a}$, $\hat{b}$,
  $\hat{c}$, and $\hat{A}$, respectively. The uncertainties are given by three standard
  deviations calculated as described in \cite[Sec. \ref{sec:crr}]{Note0}.
}

\begin{ruledtabular}
\begin{tabular}{c | r r r r}
$A$ & $\hat{a}$ & $\hat{b}$ & $\hat{c}$ & $\hat{A}$ \\
\hline
$0.6$ & $1.06\pm0.16$ & $-2.03\pm0.48$  & $1.02\pm0.24$ & $0.63\pm0.15$ \\
$0.7$ & $1.13\pm0.14$ & $-2.00\pm0.43$  & $1.01\pm0.22$ & $0.75\pm0.13$ \\
$0.8$ & $1.17\pm0.15$ & $-2.09\pm0.43$  & $1.05\pm0.22$ & $0.91\pm0.13$ \\
$0.9$ & $1.13\pm0.18$ & $-2.07\pm0.46$  & $1.04\pm0.23$ & $0.99\pm0.15$ \\
$1.0$ & $1.14\pm0.19$ & $-2.16\pm0.45$  & $1.08\pm0.22$ & $1.12\pm0.15$ \\
$1.1$ & $1.14\pm0.21$ & $-2.15\pm0.39$  & $1.08\pm0.20$ & $1.21\pm0.14$ \\
$1.2$ & $1.15\pm0.28$ & $-2.27\pm0.39$  & $1.14\pm0.20$ & $1.37\pm0.17$ \\
$1.3$ & $1.08\pm0.42$ & $-2.33\pm0.34$  & $1.16\pm0.17$ & $1.44\pm0.23$ \\
\end{tabular}
\end{ruledtabular}
\end{table}

Histograms of the time series $x(t)$ agree with Eq.~(\ref{eq:px}) for all values
of the parameter $A$ that we explored (Fig.~\ref{fig:px}). When a simulation does
not last long enough to observe sufficiently many transitions over the energy barrier $\epsilon$,
the sample of $x$ may be biased toward the Duffing potential well in which the oscillator spends
more time. The symmetry $p(-x) = p(x)$ may therefore appear imperfect in histograms
of $x$. Another issue may emerge if the time
resolution of the sample $x(t)$ is not sufficient to observe the trajectory of fast
transitions between the two wells ($-\xi < x < \xi$). In this case the histogram's
peaks overestimate the probability density at $x\approx\pm\xi$ and underestimate
it at $x\approx0$ with respect to Eq.~(\ref{eq:px}).

By using the maximum-likelihood fitting of Eq.~(\ref{eq:px}) to the time series,
we can determine the values of the parameters $\xi$ and $\sigma = A / \sqrt{a c}$.
The density $p(x)$ estimated in this way is graphically indistinguishable from the theoretical prediction
plotted in Fig.~\ref{fig:px}. Because Eq.~(\ref{eq:px}) is not sensitive to the sample
biases that are discussed above, the curve fitting of $p(x)$ yields very reliable
results.

The trajectories $x_{\pm\xi}(t)$ were selected from the time series $x(t)$ with $\varepsilon = 3 \sigma / (2 \xi)$
\footnote{
  We set $\varepsilon$ to three standard deviations of the Gaussian approximation
for Eq.~(\ref{eq:px}) at the most likely values $x=\pm\xi$
}. Profiting from the symmetry of Eq.~(\ref{eq:main}), we combined these two samples:
$\{x_\varepsilon(t)\} = \{x_{\xi}(t)\} \cup \{|x_{-\xi}(t)|\}$. As expected, the
local time autocorrelation function of $x_\varepsilon(t)$ agrees well with Eq.~(\ref{eq:crr1})
in the interval $0\le t \lesssim a^{-1}$ (Fig.~\ref{fig:crr1}). When the external
force is too weak to drive transitions between the potential wells of the Duffing
oscillator, the same theoretical expression matches perfectly the autocorrelation
function of $x(t)$.

The local time autocorrelation function of $x_\varepsilon(t)$ has an undulatory shape.
Note that Eq.~(\ref{eq:crr1}) predicts quite accurately the frequency of these undulations
even when $t \gg a^{-1}$; the discrepancy is due to their amplitude. This observation
can be explained by considering higher-order contributions $\gamma_{i=2,3...}$ \cite[Sec. \ref{sec:crr}]{Note0}.

Good estimates of the frequency $\Omega$ can be obtained by fitting Eq.~(\ref{eq:crr1})
to the time-series autocorrelations at small $t\lesssim a^{-1}$. The parameter $a$
that controls the decay of the amplitude of $\chi(t)$, however, is very sensitive
to small errors introduced by the approximate expression $\chi_1(t)$. Because the
second-order correction from Eq.~(\ref{eq:Volterra}) already leads to an unwieldy
expression for $\chi(t)$, we propose instead a phenomenological equation motivated by the general form of higher-order Volterra kernels \cite[Sec. \ref{sec:crr}]{Note0}:
\begin{multline}\label{eq:apx}
 \chi(t) \approx \hat\chi(t) = (1-\alpha) \exp(-a\,|t|)\\
  + \exp(-a\,|t|/2) [\alpha\cos(\Omega\,|t|) + \beta\sin(\Omega\,|t|)],
\end{multline}
in which $\alpha$ and $\beta$ are unknown parameters, whereas the explicit expression
can be substituted for $\Omega=\sqrt{-2 b-a^2/4}$.

Curve fitting of Eq.~(\ref{eq:apx}) on the interval $0\le t\lesssim a^{-1}$ with
four unknown parameters---$a$, $b$, $\alpha$, and $\beta$---yields quite accurate
values for $a$ and $b$. Technical details of this procedure are available in \cite[Sec. \ref{sec:crr}]{Note0}.
The numerical values of $c$ and $A$ can be found from the estimates of $a$, $b$,
$\xi$, and $\sigma$ (Table~\ref{tbl:fit}).

As shown above, the values of all four parameters of Eq.~(\ref{eq:main})---$a$, $b$,
$c$, and $A$---can be inferred from the bistable time series $x(t)$. Our approach
is limited to moderate noise intensities $A \sim \sqrt{a \epsilon}$, for which the
first term of the Volterra expansion provides a tenable approximation (Table~\ref{tbl:fit}).
In a sense this method extends the statistical techniques that were developed for
the harmonic oscillator driven by white noise \cite{BelousovCohen,Belousov2017}.

A precise quantitative description of stochastic nonlinear systems is necessary to
advance our understanding of complex behaviors observed in physics, engineering,
and biology. The Volterra expansion offers important insights into the statistical
theory of such systems. In a future communication we will present analysis of another
classical model, the Van der Pol oscillator. The convergence issues of Eq.~(\ref{eq:Volterra})
may also stimulate interest in the Wiener theory of orthogonal functional series
\cite[Chapter 9]{Schetzen}. This development might even lead to more advanced theoretical
results for the time autocorrelation function of a nonlinear oscillator. As we demonstrated
above, the analysis of autocorrelations may provide a reliable estimation of a model's
parameter values from experimental measurements.

% Separate supplemental material
\onecolumngrid
% \clearpage
\vskip0.5in
\begin{center}
  \textbf{\large Supplemental Material}
\end{center}
\twocolumngrid

\section{\label{sec:mono} The monostable Duffing oscillator}
In this section we review the Volterra-series representation of solutions for the
Duffing Eq.~(\ref{eq:main}) in the monostable regime of oscillations ($b\ge0$) \cite{Khan2001}.
Because we study a stationary problem endowed with a time-invariant probability density,
Eq.~(\ref{eq:Volterra}) can be recast as
\begin{multline}\label{eq:recast}
  x(t|f) = x_0 + \int_{-\infty}^\infty dt_1\, g_1(t-t_1) f(t_1)\\
    + \iint_{-\infty}^\infty dt_1 dt_2\, g_2(t-t_1,t-t_2) f(t_1) f(t_2) + ...\\
    = x_0(t) + \gamma_1(t) + \gamma_2(t) + ...,
\end{multline}
in which the integration limits are extended to infinities by invoking the causality
of the kernels $g_i(..., t_j,...)=0$ when $t_j\le0$, and by assuming the initial
condition $x(-\infty)=x_0\equiv\const$. We will omit indication of the infinite integration
limits in the following.

The trivial equilibrium of Eq.~(\ref{eq:zero}) is the most convenient expansion point
$x_0\equiv0$ for Eq.~(\ref{eq:recast}). Equations~(\ref{eq:one}), (\ref{eq:two}),
\textit{etc.} then become
\begin{eqnarray}
  \label{eq:first}
  \ddot\gamma_1 + a \dot\gamma_1 + b \gamma_1 &=& f ,
  \\\label{eq:second}
  \ddot\gamma_2 + a \dot\gamma_2 + b \gamma_2 &=& 0,
  \\\label{eq:third}
  \ddot\gamma_3 + a \dot\gamma_3 + b \gamma_3 &=& -c\gamma_1^3,
  \\&...&\nonumber
\end{eqnarray}
The above equations describe essentially the same harmonic oscillator
\begin{equation}\label{eq:mono}
  \ddot\gamma_i + a \dot\gamma_i + b \gamma_i = f_i
\end{equation}
subject to different forcing terms $f_i = f,0,-c\gamma_1^3...$ The left-hand side
of Eq.~(\ref{eq:mono}) corresponds to the linearized Duffing system that can be obtained
by neglecting the nonlinear cubic term in Eq.~(\ref{eq:main}).

By definition the Green function of Eq.~(\ref{eq:mono}) is the linear Volterra kernel
$g_1$:
\begin{align}\label{eq:Green}
  \gamma_1(t) =& \int ds\, g_1(t-s) f(s),\\
  g_1(t) =& \frac{2 H(t)}{\sqrt{4 b - a^2}}
    \exp\left(-\frac{a t}{2}\right)
    \sin\left(\frac{\sqrt{4 b - a^2} t}{2}\right),
\end{align}
in which $H(t)$ is the Heaviside step function. We immediately see that $\gamma_2(t)\equiv0$,
which implies that the quadratic Volterra kernel vanishes identically. The cubic
term $\gamma_3(t)$ is given by
$$ \gamma_3(t) = -c \int ds\, g_1(t-s) \gamma_1(s)^3. $$

Owing to our choice of $x_0=0$, all the Volterra kernels of even order ($g_i,i=2,4...$)
vanish. This result reflects the symmetry of the Duffing Eq.~(\ref{eq:main}). The
even-order kernels give rise to the statistical moments $\avg{x^j},j=2,4\dots$, which
must also vanish in the symmetric system with a time-invariant probability density
$p(x)=p(-x)$. The monostable Duffing oscillator may therefore be described quite
accurately by linear-response theory. Indeed, the error of such a representation
is of the order $f^3$:
\begin{equation}\label{eq:xmono}
  x(t) = \gamma_1(t) + \mathcal{O}(f^3).
\end{equation}

The stationary solutions of Eq.~(\ref{eq:mono}) are described by a bounded Green
function ($\int dt\, |g_1(t)|<\infty$) when $a > 0$ and $b > 0$. Because the higher-order
kernels have the same property, the series Eq.~(\ref{eq:recast}) converge with $x_0=0$
\cite{Ku1966}. One may then estimate statistical properties of the stationary solutions
from Eq.~(\ref{eq:xmono}).

\section{\label{sec:bi} The bistable Duffing oscillator}
The Volterra kernels found from Eqs.~(\ref{eq:first}), (\ref{eq:second}), \textit{etc.}
diverge for $t\to\infty$ when $a < 0$ or $b < 0$. The latter case corresponds to
the bistable regime of the Duffing oscillator. Equation~(\ref{eq:mono}) then describes
an unstable system and the statistical properties of the stationary solution $x(t)$ can
no longer be calculated from Eq.~(\ref{eq:xmono}). Because the ensuing Volterra kernels
are unbounded, even the existence of the expansion Eq.~(\ref{eq:recast}) can not
be ascertained.

Equation~(\ref{eq:first}) fails when $b<0$, because it represents a linearization
of Eq.~(\ref{eq:main}) around an unstable equilibrium point---the local maximum of
the Duffing potential at $x = 0$. One may however construct a convergent series
Eq.~(\ref{eq:recast}) at the minima of the potential wells $x_0=\pm\xi$. Instead
of Eq.~(\ref{eq:first})-(\ref{eq:third}) we then obtain [cf. Eq.~(\ref{eq:hrm})]
\begin{eqnarray}
  \label{eq:gamma1}
  \ddot{\gamma}_1 + a \gamma_1 - 2 b \gamma_1 &=& f,
  \\\label{eq:gamma2}
  \ddot\gamma_2 + a \dot\gamma_2 -2 b\gamma_2 &=& \mp 3 c \xi \gamma_1^2,
  \\\label{eq:gamma3}
  \ddot\gamma_3 + a \dot\gamma_3 - 2 b \gamma_3 &=& -c (\gamma_1^3 \pm 6 \xi \gamma_1 \gamma_2).
\end{eqnarray}
The linear Volterra kernel is given by the Green function of Eq.~(\ref{eq:gamma1})
(Sec.~\ref{sec:mono}):
\begin{align}\label{eq:Green1}
  \gamma_1(t) =& \int ds\, g_1(t-s) f(s),\\
  g_1(t) =& \frac{H(t)}{\Omega}
    \exp\left(-\frac{a t}{2}\right)
    \sin\left(\Omega t\right).
\end{align}
From Eqs.~(\ref{eq:gamma2}) and (\ref{eq:gamma3}) we also find the quadratic and
cubic response terms in the form
\begin{align}\label{eq:cnv2}
  \gamma_2(t) =& \mp 3 c\xi  \int ds\, g_1(t-s) \gamma_1^2
    = \mp 3 c\xi \int ds\, g_1(t-s)
      \nonumber\\&\times\iint ds_1\,ds_2\,g_1(s-s_1) g_1(s-s_2) f(s_1) f(s_2),
  \\\label{eq:cnv3}
  \gamma_3(t) =& -c \int ds\, g_1(t-s) [\gamma_1(s)^3 \pm 6 \xi \gamma_1(s) \gamma_2(s)]
    \nonumber\\=& \int ds\, g_1(t-s) \iiint ds_1\,ds_2\,ds_3\, g_1(s-s_1)
    \nonumber\\&\times [
      g_1(s-s_2) g_1(s-s_3) \pm 6\xi g_2(s-s_2, s-s_3)
    ]
    \nonumber\\&\times f(s_1) f(s_2) f(s_3).
\end{align}

To extract the quadratic and cubic kernels $g_{i=2,3}$ from the above equations we
rely on a simplified growing-exponential approach \cite[Sec. 3.5]{Rugh}. We use a
substitution rule for the product of the forcing terms in the form
\begin{equation}\label{eq:sub}
  f(s_1) f(s_2)\cdots \to\exp(-\imi\omega_1 s_1 - \imi\omega_2 s_2-\dots).
\end{equation}
The results that are obtained for arbitrary $\omega_1$ and $\omega_2$ hold also in
the special case $\omega_1=\omega_2$. Like the sum of growing exponentials \cite[Sec. 3.5]{Rugh},
our approach also renders the symmetric form of the Volterra kernels [$g_i(...s_j, s_k...) = g_i(...s_k,s_j...)$].
In general we have
\begin{multline}\label{eq:gen}
  \gamma_i(t) = \int\cdots\int ds_1 \cdots ds_i\, g_i(t-s_1,\dots,t-s_i)
    \\\times\exp(-\imi \omega_1 s_1 \dots -\imi\omega_i s_i)
    \\= G_i(\omega_1,\dots,\omega_i) \exp[\imi t (\omega_1 +\dots+ \omega_i)],
\end{multline}
in which $G_i$ is the Fourier transform of the kernel $g_i$.

By substituting Eq.~(\ref{eq:sub}) into (\ref{eq:cnv2}) we obtain
\begin{multline}\label{eq:sub2}
  \gamma_2(t) = \mp 3 c\xi\,G_1(\omega_1) G_1(\omega_2) G_1(\omega_1+\omega_2)
    \\\times\exp[\imi t (\omega_1 + \omega_2)],
\end{multline}
and by comparing the above equation with Eq.~(\ref{eq:gen}) we identify the Fourier
image of the quadratic kernel
\begin{equation}\label{eq:G2}
  G_2(\omega_1, \omega_2) = \mp 3 c\xi\,G_1(\omega_1) G_1(\omega_2) G_1(\omega_1+\omega_2).
\end{equation}
We likewise find the Fourier transform of the cubic kernel
\begin{multline}\label{eq:G3}
  G_3(\omega_1, \omega_2, \omega_3) =
    - c G_1(\omega_1 + \omega_2 + \omega_3)
    \\\times G_1(\omega_1) G_1(\omega_2) G_1(\omega_3) [1 - 18 b G_1(\omega_2 + \omega_3)].
\end{multline}
Because the expressions for $g_2$ and $g_3$ are unwieldy, further calculations are
more convenient in Fourier space.

The terms of even orders do not vanish in the Volterra series about $x_0=\pm\xi$,
because the potential wells surrounding these points are asymmetric. The global symmetry
of the Duffing potential $U(x)=U(-x)$ ensures, however, that the even-order Volterra
kernels for $x_0=\pm\xi$ have opposite signs, whereas the odd-order kernels coincide
[\textit{cf.} Eqs.~(\ref{eq:G2}) and (\ref{eq:G3})].

\section{\label{sec:crr} Local autocorrelations of the bistable Duffing oscillator}

\begin{figure*}
  \begin{minipage}[t]{\columnwidth}
    \includegraphics[width=\columnwidth]{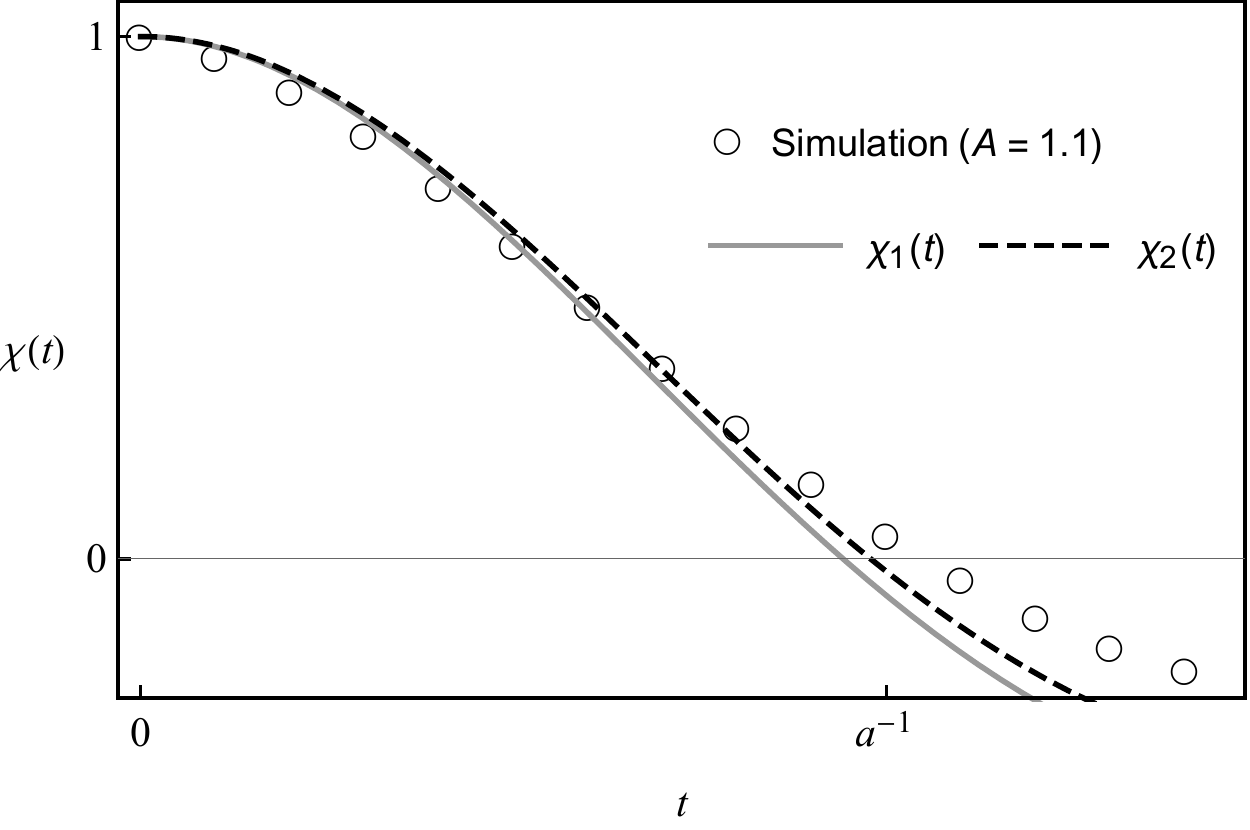}
    \caption{\label{fig:crr2}
      Local time autocorrelation function $\chi(t)$ of the time series $x_\varepsilon(t)$:
      comparison of theoretical predictions $\chi_1(t)$ and $\chi_2(t)$ [Eqs.~(\ref{eq:crr1})
      and (\ref{eq:crr2})] with the simulation results with $A = 1.1$. The corrections
      that are introduced in $\chi_2(t)$ by including the second-order response term
      in Eq.~(\ref{eq:recast}) provide a subtle improvement over $\chi_1(t)$.
    }
  \end{minipage}
  \hfill
  \begin{minipage}[t]{\columnwidth}
    \includegraphics[width=\columnwidth]{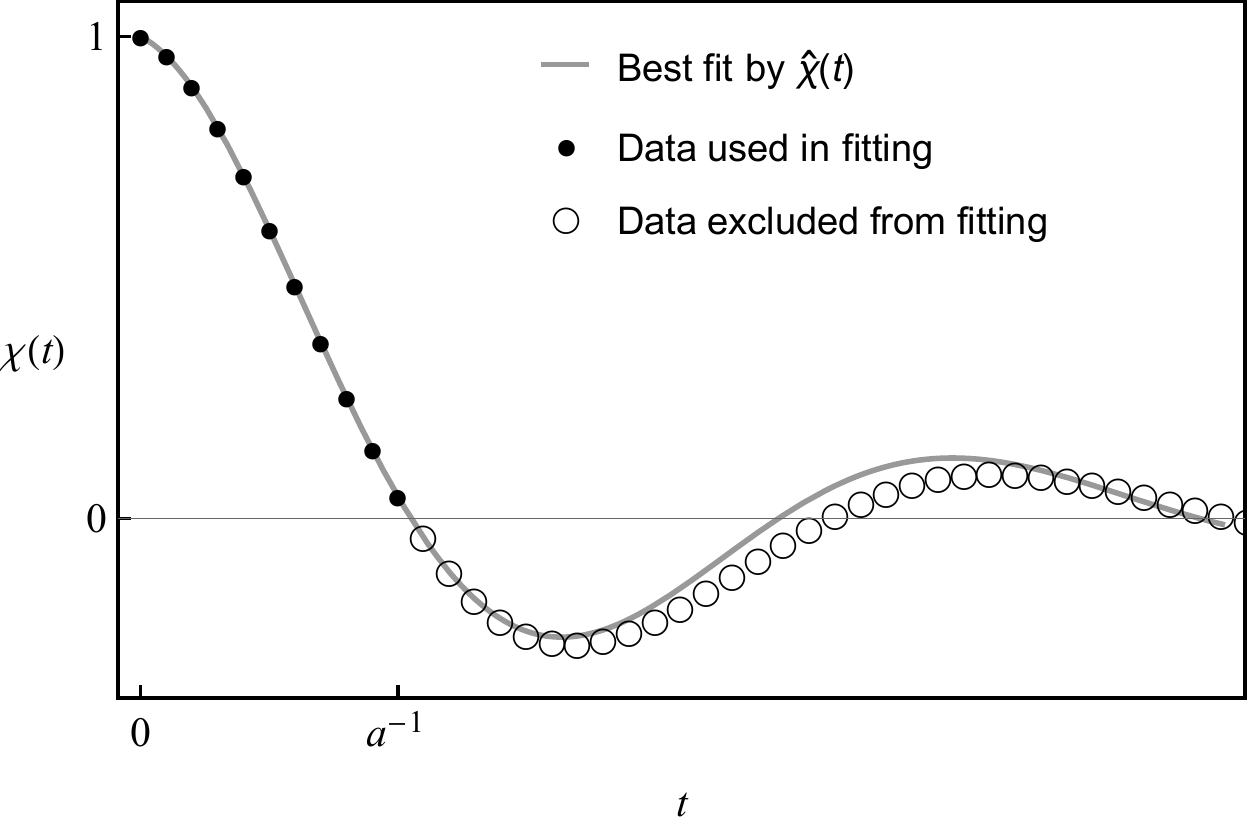}
    \caption{\label{fig:fit}
      Curve fitting of $\hat\chi(t)$ to the time autocorrelation function of $x_\varepsilon(t)$
      observed in simulations with $A=1.1$. The expression optimized in the interval
      $0\le t \lesssim a^{-1}$ extrapolates well up to $t \lesssim 2 a^{-1}$.
    }
  \end{minipage}
\end{figure*}

Equation~(\ref{eq:crr1}), which approximates the autocorrelation function $\chi(t)$,
has been derived from the Volterra expansion truncated at the linear term $\gamma_1(t)$
given by Eq.~(\ref{eq:Green1}). To find a second-order correction, one may include
a quadratic contribution in $x(t) \approx \gamma_1(t) + \gamma_2(t)$:
\begin{equation}\label{eq:crr2}
  \chi(t) \approx \chi_2(t) = \frac{\theta_1(t) + \theta_2(t)}{\theta_1(0) + \theta_2(0)},
\end{equation}
in which
\begin{align}
  \theta_1(t) =& -\frac{A^2}{4 a b} \chi_1(t),\\
  \theta_2(t) =& \avg{\gamma_2(0)\gamma_2(t)} - \avg{\gamma_2}^2.
\end{align}
From \cite[Eq.~(11.3-14)]{Schetzen} one can find the Fourier transform of $\theta_2$
\begin{multline}\label{eq:Theta2}
  \Theta_2(\omega) = \frac{A^4}{\pi} \int d\lambda\, |G_2(\lambda,\omega-\lambda)|^2
    \\= -\frac{9 b c A^4}{\pi} |G_1(\omega)|^2
    \int d\lambda\, |G_1(\omega)|^2\,|G_1(\omega-\lambda)|^2.
\end{multline}
Note that
$$ \theta_1(t) = \int \frac{d\omega}{2\pi A^2} |G_1(\omega)|^2 \exp(\imi\omega t). $$
By virtue of the convolution theorem, the inverse Fourier transform of Eq.~(\ref{eq:Theta2})
then yields
\begin{widetext}
\begin{multline}\label{eq:theta2}
  \theta_2(t) \propto \int ds\, \chi_1(t-s) \chi_1(s)^2
    = \const_0 \exp(-a\,|t|)
    + \exp(-a\,|t|/2)\,[\;
      \const_1 \sin(\Omega\,|t|) + \const_2 \cos(\Omega\,|t|)
    \;]\\
    + \exp(-a\,|t|/2)\,[\;
      \const_3 \sin(2 \Omega\,|t|)
      + \const_4 \cos(2 \Omega\,|t|)\;
    ],
\end{multline}
\end{widetext}
in which the unwieldy constants $\const_{i=0,1,2,3,4}$ are not spelled out for clarity.
These coefficients, which can be readily found with the help of a symbolic computational
software \cite{Mathematica}, depend in a complicated manner on all four parameters
of the Duffing oscillator.

The complete expression of $\chi_2(t)$ reveals higher-order harmonics $j \Omega,\,j=2,3...$
in the autocorrelation function. Additional oscillations at these frequencies are
commensurate with the undulations of $\chi_1(t)$. For this reason, as noted in the
main text, Eq.~(\ref{eq:apx}) predicts correctly the undulatory component of $\chi(t)$.

Due to the complexity of the explicit expression for $\chi_2(t)$, using Eq.~(\ref{eq:crr2})
in calculations and curve fitting is problematic. The improvement that is achieved
over Eq.~(\ref{eq:crr1}) is also modest (Fig.~\ref{fig:crr2}). Higher-order expressions
that take into account more terms from Eq.~(\ref{eq:recast}), might be formidably
long. In curve fitting we therefore use a phenomenological Eq.~(\ref{eq:apx}) as
justified below.

Contemplating the development of the Volterra terms in Eqs.~(\ref{eq:gamma1})--(\ref{eq:G3}),
one may expect that $\chi(t)\approx\chi_n(t)$, calculated from Eq.~(\ref{eq:recast})
with $n$ response terms, contains only convolution and power products of $\chi_1(t)$
[\textit{cf.} Eq.~(\ref{eq:theta2})]. The resulting expression would be a composition
of time-dependent factors in the form $\propto\exp(i a t/2)$, $\propto\sin(j \Omega t)$,
$\propto\cos(k \Omega t)$ with integers $i\ge2,\,j\ge1,\,k\ge1$. Whereas we retain
the fundamental harmonic terms $\propto\sin(\Omega t)$ and $\propto\cos(\Omega t)$,
as well as the slowly decaying exponential $\propto\exp(a t)$ in $\chi_2(t)$, we
introduce unknown coefficients $\alpha$ and $\beta$ to account for higher-order corrections.
By imposing an additional constraint $\chi(0)=1$, we arrive at Eq.~(\ref{eq:apx}).

Finally, we discuss briefly the procedure of curve fitting for Eq.~(\ref{eq:apx})
(Fig~\ref{fig:fit}). By using the criterion of Lagarkov and Sergeev \cite{Lagarkov1978,BelousovCohen},
we select from the sample autocorrelation data the observations in the interval $0 \le t \le t_0$,
in which $t_0$ is the instant when the autocorrelation function reaches the value
of zero for the first time [$\chi(t_0) = 0$]. This choice of $t_0$ corresponds approximately
to the relaxation time $t_0\approx\tau$, for which Eq.~(\ref{eq:apx}) should give
accurate results. As an initial guess we recommend setting $\alpha=0$ and $\beta=0$.
Otherwise the least-square fitting of Eq.~(\ref{eq:apx}), which is a flexible expression
with four unknown parameters, may return suboptimal results.

A simple least-square fitting of the phenomenological Eq.~(\ref{eq:apx}) underestimates
standard errors of the parameter values. We report more realistic estimates, which are recalculated
by using the optimized parameter values and the uncertainty of the time autocorrelation
data \cite{Bowly1928}
$$ \Delta{\chi}\approx \sqrt{\frac{1-\chi^2}{n-2}}. $$

\section{\label{sec:sim}Simulation algorithm}
For computational experiments we convert Eq.~(\ref{eq:main}) into an equivalent two-dimensional
dynamical system $\pnt{X} = (x,y) = (x,\dot{x})$:
\begin{equation}
  \begin{cases}
    \dot{x} = y\\
    \dot{y} =  -a y - b x - c x^3 + f(t).
  \end{cases}
\end{equation}
We adopt a second-order operator-splitting approach for stochastic systems \cite[Appendix~C]{Belousov2017}
by decomposing the time-evolution operator $\oper{T}$ as
\begin{equation}\label{eq:split}
  \dot{\pnt{X}} = \oper{T} \pnt{X} = (
    \oper{T}_{y,x} + \oper{T}_{x,y} + \oper{T}_{y,y}
  ) \pnt{X},
\end{equation}
in which
\begin{align*}
  &\oper{T}_{y,x} = y \partial_x,\quad
    \oper{T}_{y,y} = - a y \partial_y,\\
  &\oper{T}_{x,y} = (f - b x - c x^3) \partial_y.
\end{align*}

The formal solution of Eq.~(\ref{eq:split}) for a time step $\Delta{t}$ is
$$ \pnt{X}(t+\Delta{t}) = \exp(\oper{T} \Delta{t}) \pnt{X}(t), $$
in which the time-evolution operator can be approximated by
\begin{multline}\label{eq:operator}
  \exp[\oper{T}\Delta{t} + O(\Delta{t}^2)] =
      \exp\left(\frac{\oper{T}_{y,x}\Delta{t}}{2}\right)
      \exp\left(\frac{\oper{T}_{y,y}\Delta{t}}{2}\right)\\
      \times\exp(\oper{T}_{x,y} \Delta{t})
      \exp\left(\frac{\oper{T}_{y,y}\Delta{t}}{2}\right)
      \exp\left(\frac{\oper{T}_{y,x}\Delta{t}}{2}\right)
    .
\end{multline}
The action of an individual operator of the form $\exp(\oper{L}\Delta{t})$ can be
inferred by solving the simplified dynamics
\begin{equation}
  \dot{\pnt{X}}(t) = \oper{L} \pnt{X}(t) \Rightarrow \pnt{X}(t+\Delta{t})
    = \exp(\oper{L}\Delta{t}) \pnt{X}(t).
\end{equation}
The composite operator~(\ref{eq:operator}) then leads to the following algorithm
for the numerical integration of Eq.~(\ref{eq:split}):
\begin{align}
  &x(t+\Delta{t}/2) = x(t) + y(t)\Delta{t} / 2,\\
  &y(t+\Delta{t}) =
    y(t)\exp(-a \Delta{t})\nonumber\\&\quad
    - \exp(-a \Delta{t}/2)[b + c x(t+\Delta{t}/2)^2] x(t+\Delta{t}/2) \Delta{t}
    \nonumber\\&\quad
    + \exp(-a \Delta{t}/2) \int_t^{t+\Delta{t}} dt f(t),\\
  &x(t+\Delta{t}) = x(t+\Delta{t}/2) + y(t+\Delta{t})\Delta{t}/2.
\end{align}

\bibliographystyle{apsrev4-1}
\bibliography{refs}

\end{document}